\documentclass[a4paper,fleqn,usenatbib]{mnras}

\usepackage{newtxtext,newtxmath}

\usepackage[T1]{fontenc}
\usepackage{ae,aecompl}

\usepackage{graphicx}	
\usepackage{amsmath}	
\usepackage{amssymb}	

\newcommand{\grp}    {${\rlap.}^{\circ}$}

\newcommand{\ltsima} {$\; \buildrel < \over \sim \;$}
\newcommand{\simlt}  {\lower.5ex\hbox{\ltsima}}            
\newcommand{\gtsima} {$\; \buildrel > \over \sim \;$}
\newcommand{\simgt}  {\lower.5ex\hbox{\gtsima}}            

\newcommand{\fgl} {4FGL J0647.7$-$4418}

\title{A blazar as the likely counterpart to 4FGL J0647.7-4418 instead of a gamma-ray binary}

\author[J. Mart\'{\i} et al.]{
Josep Mart\'{\i},$^{1}$\thanks{E-mail: jmarti@ujaen.es}
Estrella S\'anchez-Ayaso,$^{2}$
Pedro L. Luque-Escamilla,$^{3}$
Josep M. Paredes$^4$,
\newauthor{Valent\'{\i} Bosch-Ramon$^{4}$, and Robin H. D. Corbet$^{5}$}
\\
$^{1}$Departamento de F\'isica (EPSJ), Universidad de Ja\'en, Campus Las Lagunillas s/n Ed. A3, E-23071 Ja\'en, Spain\\
$^2$ Departamento de Ciencias Integradas, Facultad de Ciencias Experimentales, Campus "El Carmen", Universidad de Huelva,
 Avda. de las Fuerzas \\ Armadas s/n, E-21007  Huelva, Spain \\
$^{3}$Departamento de Ingenier\'{\i}a Mec\'anica y Minera (EPSJ), Universidad de Ja\'en, Campus Las Lagunillas s/n Ed. A3, E-23071 Ja\'en, Spain\\
$^{4}$Departament de F\'{\i}sica Qu\`antica i Astrof\'{\i}sica,  Institut de Ci\`encies del Cosmos, Universitat de Barcelona, IEEC-UB, Mart\'{\i} i Franqu\`es 1, \\ E-08028 Barcelona, Spain\\
$^{5}$University of Maryland, Baltimore County, and X-ray Astrophysics Laboratory, Code 662 NASA Goddard  Space Flight Center,  Greenbelt Rd., \\ MD 20771, USA\\
}

\date{Accepted XXX. Received YYY; in original form ZZZ}

\pubyear{2019}

\begin{document}
\label{firstpage}
\pagerange{\pageref{firstpage}--\pageref{lastpage}}
\maketitle

\begin{abstract}
The  persistent gamma-ray source 4FGL J0647.7$-$4418 is tentatively associated in the latest {\it Fermi} catalogue with the sub-dwarf O-type X-ray binary HD 49798.
However,  an AGN candidate is also mentioned as an alternative identification in updated versions of the catalogue accompanying paper.
If the first association were correct, this would add HD 49798 to the handful of currently known gamma-ray binaries and, therefore, represent a significant breakthrough not only because
of a new member addition, but also because of the apparent white dwarf companion in this system.
Despite these perspectives,
here we show that the stellar association is likely wrong and that the proposed AGN object, 
well inside the {\it Fermi} 95\% confidence ellipse, is a more conceivable counterpart candidate to the {\it Fermi} source due to its strong blazar similarities.

\end{abstract}

\begin{keywords}
gamma rays: stars -- X-rays: binaries -- white dwarfs -- stars: individual: HD49798 -- (galaxies:) BL Lacertae objects: general
\end{keywords}



\section{Introduction}
 
Gamma-ray binaries (GBs) provide one of the most extreme cases of  electromagnetic radiative output from
gravitationally bound stellar pairs, reaching not only high photon energies (HE, $\sim 1$ GeV),  but also very-high-energies (VHE, $\sim 1$ TeV) in some systems.
This is a relatively new field in gamma-ray astrophysics that has been intensively reviewed several times in the past years
\citep{2013A&ARv..21...64D, 2015CRPhy..16..661D, 2019arXiv190103624P}. 
In particular,  high-mass GBs whose optical counterpart is an early-type  (O or B) luminous star, 
with or without emission lines from a circumstellar envelope, are the most distinctive members of the GB family.
The taxonomy of stellar gamma-ray sources also includes a variety of other related systems,
such as microquasars, colliding wind binaries, novae, and transitional millisecond pulsars \citep{2019RLSFN.tmp...11P},  
but not often as luminous and persistent as the 'classical' GBs referred above.
Room for other kinds of unexpected gamma-ray stellar emitters could also await discovery.

The nature of the unseen companion in the most common case of 
 high-mass GBs is still under an animated debate. The exception is in  the case where pulsations have been detected and attributed to a neutron star
whose rotational energy is suspected to act as the ultimate power source.
High-mass GBs typically display both HE and VHE emission modulated with the system orbital, and sometimes also super-orbital, period(s).
Therefore,  the physical conditions for acceleration of particles and different emission 
processes repeat in a predictable way that facilitates their study.  
The wide attention raised by this special kind of systems is in contrast with the reduced number currently being known
as only $\sim 10$ sources have been detected so far. 
Moreover, the perspectives for new Milky Way discoveries do not appear to be very optimistic based on a population
synthesis analysis that estimated a total number of only $\sim 10^2$ systems in our Galaxy \citep{2017A&A...608A..59D}.

The GB scarcity contrasts with the dominant abundance of blazar-type sources across the whole sky in the present and past {\it Fermi} catalogue series.
The latest of these, nicknamed as 4FGL, was released 
after 8 year of data collection with the Large Area Telescope (LAT) on board NASA's {\it Fermi} space observatory \citep{2019arXiv190210045T}.
Remarkably, this catalogue proposes the association of  one of its persistent sources, namely \fgl,  with the already known X-ray binary 1WGA J0648.0$-$4419 \citep{2006A&A...455.1165L}.
This fact immediately prompted us to explore in more detail the chances of a real association with the gamma-ray source as suggested by the {\it Fermi} catalogue itself.
Several of the currently recognized GBs were previously known as common X-ray binaries and we wondered if HD 49798 could follow the same path.
A real association would not only mean incremental progress in the field, but also a class-broadening of the GB family since its compact object is more consistent with 
a spun-up white dwarf rather than a neutron star. 

At this point we must indicate that, starting with the $3^{rd}$ draft version of the 4FGL accompanying paper, posted by the {\it Fermi} collaboration, 
 an important information  with respect to
this binary system was added. The {\it Fermi} team noticed that the binary association probability (85\%) was barely larger than that of a nearby
blazar candidate (80\%) not considered before.
Despite this, the latest electronic version of the 4FGL catalogue still lists the binary association as the only one
and \fgl\ remains classified as a high mass binary. In this paper, we provide significant evidence indicating that this classification is likely wrong and 
in need of revision.

\section{The GB suspect}

HD 49798 is a bright ($V=8.27$), early-type star HD 49798
located well below the Galactic plane ($l^{II} =$253\grp 71, $b^{II}=-$19\grp 14), 
and at a distance of  $508 \pm 16$ pc  
according to the {\it Gaia} Data Release 2  \citep{2018A&A...616A...1G}.
Its peculiar nature was originally pointed out by the renowned Jascheck spectroscopists, who
recognized it as a hot sub-dwarf  of O6 spectral type (sdO6, luminosity class VI)
using plates taken at the Bosque Alegre station of C\'ordoba Observatory in Argentina \citep{1963PASP...75..365J}.
Early-type, hot subdwarfs are believed to originate when a red giant star gets rid of its outer layers
of hydrogen before the core starts to fuse helium. 

The Jaschecks work was followed by a spectroscopic
orbital solution that yielded an orbital period of 1.55 d and mass function of about 0.27 $M_{\odot}$ \citep{1970MNRAS.150..215T},
values later confirmed using {\it IUE} spectra \citep{1994Obs...114...41S}. Hints about the nature of the HD 49798 companion emerged 
finally thanks to {\it ROSAT} X-ray observations \citep{1997ApJ...474L..53I}, that revealed the existence of a pulsation period of  $P=13.2$ s attributed to either a white dwarf (WD)
or a magnetic neutron star (NS). The discovery with XMM-$Newton$ of X-ray eclipses allowed to highly constrain the system orbital inclination \citep{2011ApJ...737...51M}.
These authors also combined this information with the X-ray pulse delays and the optical mass function.
In this way, it was possible to fully solve for the mass of the sub-dwarf  star ($M_* =1.50$ $M_{\odot}$) and that of the compact object ($M_X =1.28$ $M_{\odot}$).
With this information in hand, the classification of HD 49798 as a high-mass X-ray binary becomes mostly based on the spectral type earliness 
of the donor star rather than its mass.

A latter X-ray analysis by
\citet{2016MNRAS.458.3523M} remarkably revealed a steady spin-up of the pulsed emission of $\dot{P} = -2.15 \times 10^{-15}$ s s$^{-1}$,
initially attributed to accretion of the wind from the sub-dwarf companion onto a NS. However, the observed spin-up stability over $~20$ years is very unusual in NS X-ray binaries.
Recently, this effect has been reinterpreted in 
a consistent WD context  under contraction because of its young age  ($\sim 2$ Myr),  with  the current radius estimate being $R_X \simeq 3340$ km
\citep{2018MNRAS.474.2750P}.
The spin-up contribution due to stellar wind accretion from the sub-dwarf, with mass loss $\dot{M}_w = 3 \times 10^{-9}$ $M_{\odot}$ yr$^{-1}$
and terminal velocity $v_{\infty} \simeq 1350$ km s$^{-1}$ \citep{2010Ap&SS.329..151H}, is found to be negligible in this case.
Independently of the compact object nature, the physical properties of the HD 49798 system strongly depart from a typical high mass X-ray binary.
How HD 49798 could fit into the GB family, where no sub-dwarfs are found and no clear scenario for gamma-ray emission is available, started to be challenging.

\section{Re-analysis of {\it Fermi}-LAT data}

 \begin{figure}
\includegraphics[width=8.0cm]{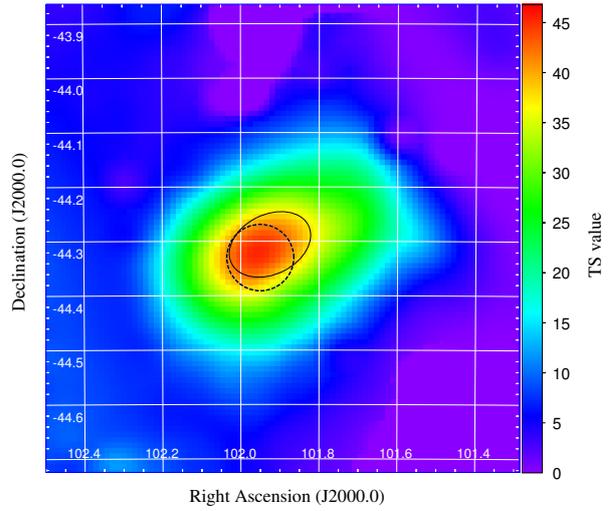}
\caption{   \label{TSmap} TS  map ($1^{\circ} \times 1^{\circ}$) of the 4FGL J0647.7$-$4418 region resulting from our own analysis of {\it Fermi}-LAT data.
The black dashed circle represents the improved 95\% confidence level gamma-ray source location.
The black ellipse corresponds to the same confidence level location according to the 4FGL catalogue.
}
\end{figure}

Our first step  was to perform a
brief re-analysis of the {\it Fermi}-LAT Pass 8 database towards HD~49798, which included 
data  from August 4, 2008 (MET 239557417) until July 1, 2019 (MET 583643061).
The immediate purpose was to
explore if some improvement with respect to the 4FGL  catalogue was achievable, specially  looking forward to a more accurate gamma-ray source location.
We selected the front and back converted photons at energies from 100 MeV to 1 TeV within a 5$^{\circ}$ $\times$ 5$^{\circ}$ square region centered at the position of the source, considered as the region of interest (ROI).
In order to exclude time periods in which some spacecraft event had affected the data quality, we used the source event class recommended for individual source analysis using the expression (DATA$\_$QUAL$>$0)\&\&(LAT$\_$CONFIG $==$1).
Moreover, only events with zenith angles lower than $90^{\circ}$  were taken into account to reduce the background contamination from the Earth's albedo.
We used the current Fermitools from conda distribution and the latest version of the instrument response functions (IRFs) P8R3\_SOURCE\_V2 
to process our data. 
All steps included in the Fermitools tutorial\footnote{https://fermi.gsfc.nasa.gov/ssc/data/analysis/scitools/\\ likelihood\_tutorial.html} for unbinned likelihood analysis
were carefully followed.
We also ran the make4FGLxml.py script\footnote{https://fermi.gsfc.nasa.gov/ssc/data/analysis/user/
make4FGLxml.py} within the ROI, employing the 4FGL catalogue and  
making use of the Galactic diffuse model gll\_iem\_v07.fits and isotropic emission model iso\_P8R3\_SOURCE\_V2\_v1.txt\footnote{https://fermi.gsfc.nasa.gov/ssc/data/access/lat/BackgroundModels.html}. 

As a result of our re-analysis we obtained   a detection very compatible with the one provided by the 4FGL catalogue, with just
a slightly fainter integral photon flux of $(2.0\pm 0.4) \times 10^{-10}$ photon cm$^{-2}$ s$^{-1}$ from 1 to 100 GeV. The resulting spectrum was
adequately described by a simple power-law type $dN/dE=N_0(E/E_0)^{-\gamma}$ with index $\gamma=2.2\pm 0.2$.
The corresponding Test Statistic (TS) value was 47, which translated into a detection significance of about 7.
This confirmed the reality of \fgl\ with two additional years of data not included in the 4FGL catalogue.

 The resulting TS map is presented in Fig. \ref{TSmap}. Our best estimate for the  J2000.0 source position
  (Right Ascension = 101\grp 9500; Declination = $-44$\grp 3300), with error radius 0\grp 06 at the 95\% confidence level,
was finally derived from it by using the task
 {\tt gtfindsource} and the {\tt plot\_tsmap.py} script\footnote{http://grbworkshop.wikidot.com/s6-fermi-lat-analysis-hands-on}.

\section{Looking for the orbital period signature in {\it Fermi} data}

\begin{figure}
\begin{center}
\includegraphics[width=6.5cm, angle=-90]{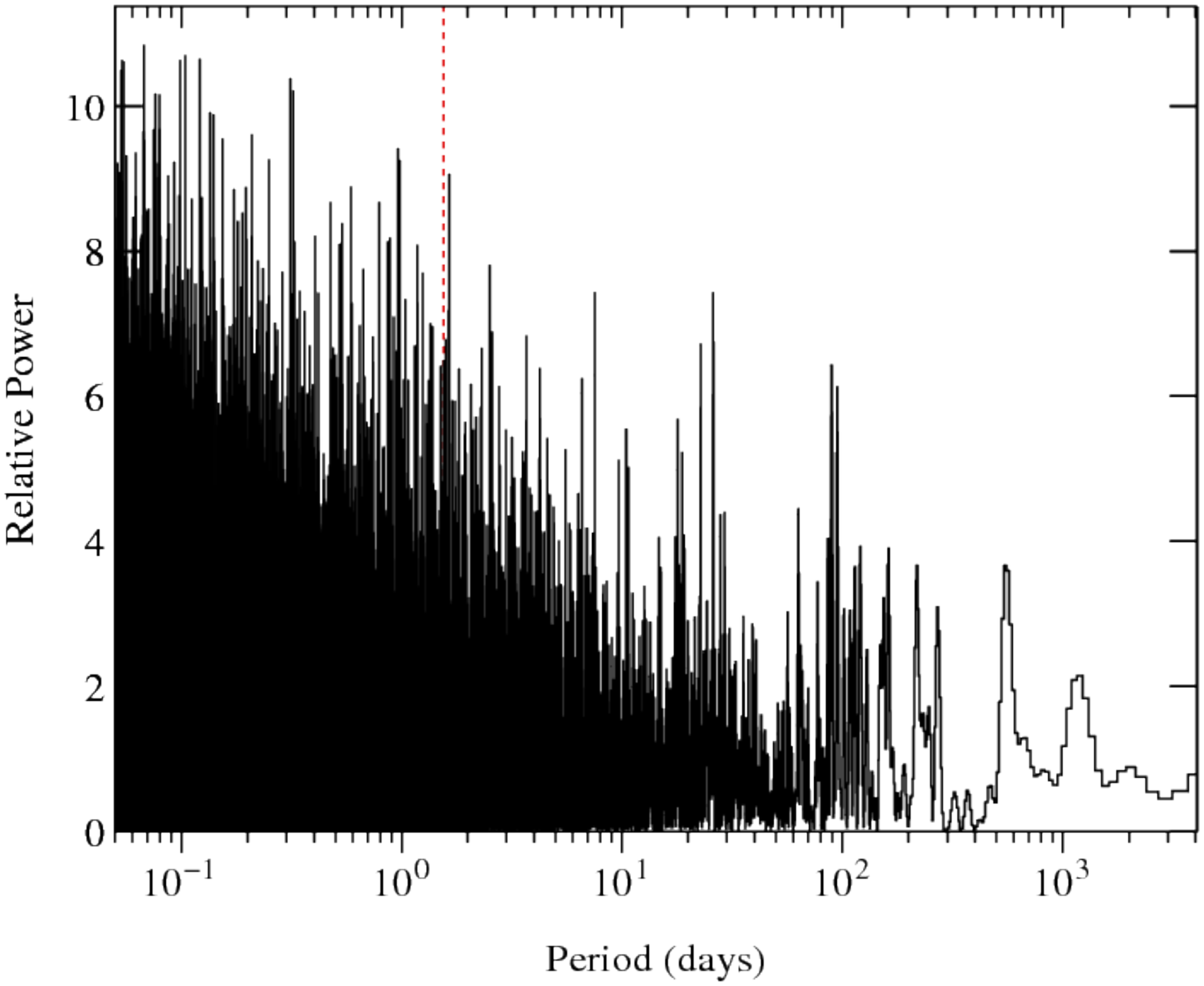}
\end{center}
\caption{Power spectrum of the {\it Fermi}-LAT light curve computed as described in the text.
The dashed red line indicates the location of the most accurate orbital period of
1.547666 d  \citep{2016MNRAS.458.3523M}  where no relevant power is seen.
\label{power}
 }
\end{figure}

The light curve of {\it Fermi}-LAT data was also extracted around the HD 49798 position in order to search its power spectrum for the known orbital period of the system.
We used data covering the interval from MJD 54682 to 58745 with a time resolution of 500 s.
The technique of probability-weighted aperture photometry was used for this purpose.
 This is the same technique that led to the most recent discoveries of the
GBs LMC P3 \citep{2016ApJ...829..105C} and 4FGL J1405.1-6119   \citep{2019ApJ...884...93C}. 
The resulting power spectrum is shown in Fig. \ref{power}. When calculating it,
weights were given according to  the relative exposure of each data point. 

The power spectrum displayed in Fig. \ref{power} does not show any significant power component matching the 
HD 49798 orbital cycle. This negative result is relevant since one might expect to see some signal at the orbital period value if the binary system was the true source of HE photons.

\section{Multi-wavelength search for alternative counterparts}

\begin{table*}
\small
\caption{X-ray sources consistent with 4FGL J0647.7$-$4418\label{xraysour}}
\begin{tabular}{cccccccccc}
\hline
              & \multicolumn{3}{c}{Source location} &  \multicolumn{3}{c}{Flux ($10^{-14}$ erg cm$^{-2}$s$^{-1}$)} & Hardness   & Optical  \\ 
  3XMM  &    R. A. &  Dec. &  Pos. Err & $F_{\rm total}$ & $F_{\rm soft}$ & $F_{\rm hard}$ & ratio & counterpart  & \\
              &  $(^{\circ})$ &  $(^{\circ})$ &  $(^{\prime\prime})$ & 0.5-12 keV & 0.5-2 keV & 2-12 keV & $\frac{F_{\rm hard}-F{\rm soft}}{F_{\rm total}}$ &  \\
\hline
J064747.1$-$441950 & 101.94638  & -44.33063  & 0.11   & 15.6 &   5.0  & 10.6 & +0.36 & AGN$^a$  \\
J064759.5$-$441941 & 101.99824    & -44.32828 & 0.09 & 18.6 & 16.8  &  1.7  & $-0.81$ & Star$^{b,c}$  \\
J064804.6$-$441858$^{d}$ & 102.01953  & -44.31623 & 0.09 &  13.7 &  5.3 &  8.4 &  +0.23  & HD 49798  \\
\hline
\end{tabular}
~\\
$^a$ Candidate AGN classification \citep{2012ApJ...756...27L}. 
$^b$ Candidate star classification \citep{2012ApJ...756...27L}. 
$^c$ Significant parallax and proper motion available \citep{2018A&A...616A...1G}.
$^d$ Also known as RX J0648.0$-$4418.
\end{table*}

\begin{figure}
\begin{center}
\includegraphics[width=\hsize]{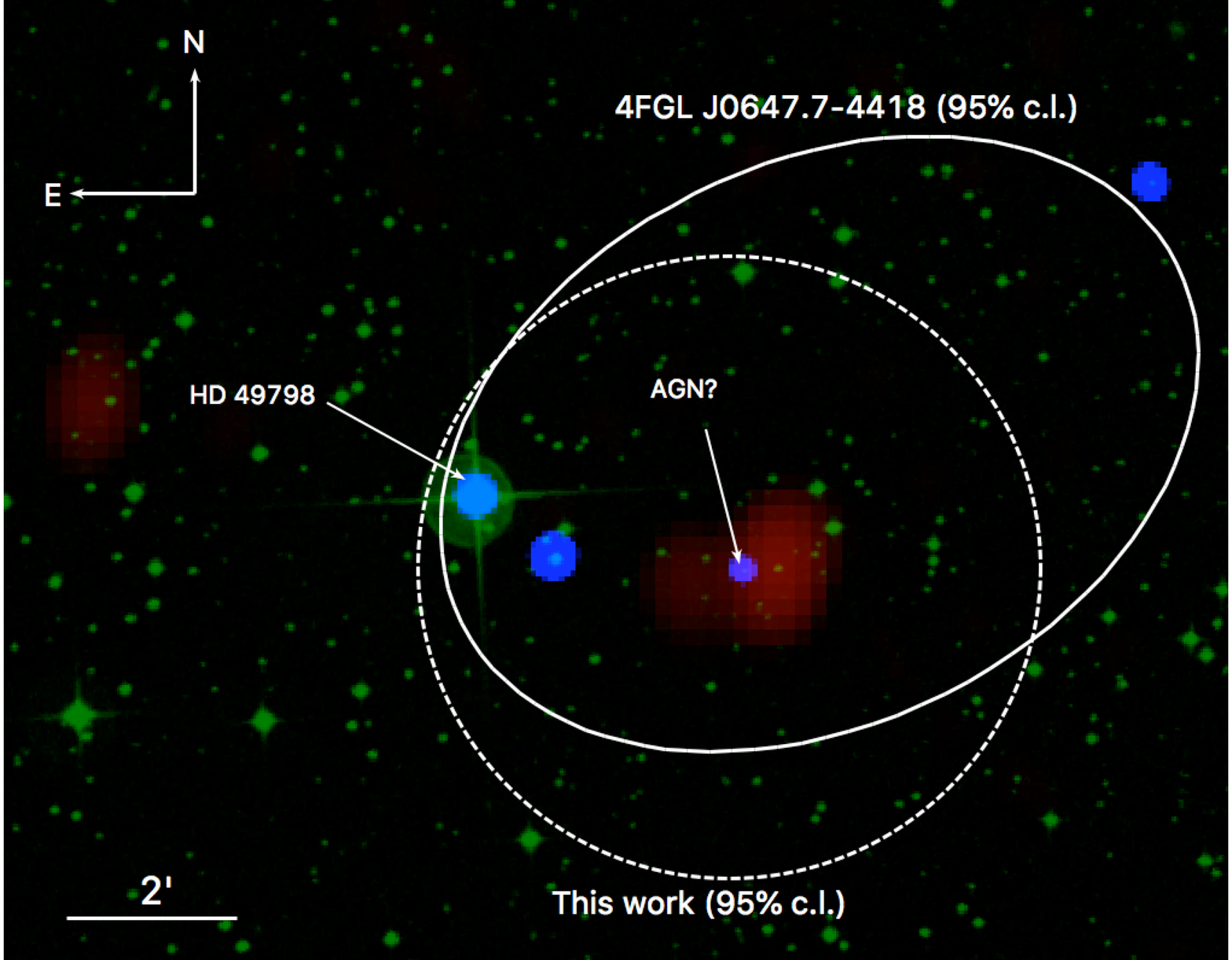}
\end{center}
\caption{Tri-chromatic view of the region of the \fgl\ confidence ellipse at the 95\% level where the red, green and blue layers
correspond to radio, optical and X-ray emission downloaded from the SUMSS 843 MHz, DSS2 blue plate and XMM-{\it Newton} archives, respectively.
The improved 95\% confidence position resulting from our own analysis of {\it Fermi} data  is also included as a dashed circle.
The location of  HD 49798 and the suspected AGN source are arrowed on the image.   \label{tricom}
 }
\end{figure}

In Fig. \ref{tricom}, a tri-chromatic view of the 4FGL J0647.7$-$4418/HD49798 field is shown with the 95\% confidence regions of 
4FGL J0647.7$-$4418 from both the {\it Fermi}-LAT  catalogue and our own work over-plotted onto it.
The blue layer corresponds to X-ray emission as observed with the  XMM-{\it Newton} observatory where, according to the 
3XMM-DR8 catalog release\footnote{{\tt  http://xmm-catalog.irap.omp.eu}} \citep{2017ApJ...839..125Z},
HD 49798 agrees within $\sim 0.1$  arcsecond with the bright X-ray source 3XMM J064804.6$-$441858. 
The green layer represents the optical appearance of the field as retrieved from 
the digitized version of the second Palomar Observatory Sky Survey (POSS II),
with the blue plate being used \citep{1991PASP..103..661R}. 
Finally, the red layer displays radio emission at 843 MHz
from the Sydney University Molonglo Sky Survey (SUMSS) \citep{2003MNRAS.342.1117M}.

Table \ref{xraysour} contains the relevant information for all the three X-ray sources consistent with the {\it Fermi} source sorted by right ascension.
The first of them is 3XMM J064747.1$-$441950, which has been proposed as a possible Active Galactic Nucleus (AGN) in \citet{2012ApJ...756...27L} and will be considered in more detail below.
The second one is 3XMM J064759.5$-$441941, which agrees  within $0.2$ arcsecond with an anonymous stellar object.
According to the {\it Gaia} Data Release 2 \citep{2018A&A...616A...1G}, both a high parallax (10.74 milli-arcsecond) and
proper motion (10.4 milli-arcsecond year$^{-1}$) are reported for this object, therefore ensuring its stellar nature.
 Assuming negligible reddening because of its proximity, the observed brightness level ($G=12.88$)
and low effective temperature ($T_{\rm eff} = 4100$ K) points to a main sequence star of M-type with chromospheric activity unlikely to produce detectable gamma-rays. 
The third and last source is 3XMM 064804.6$-$441858, the previously mentioned X-ray counterpart of HD 49798 that motivated our work.

\section{A blazar AGN counterpart}

\begin{figure}
\begin{center}
\includegraphics[scale=0.4]{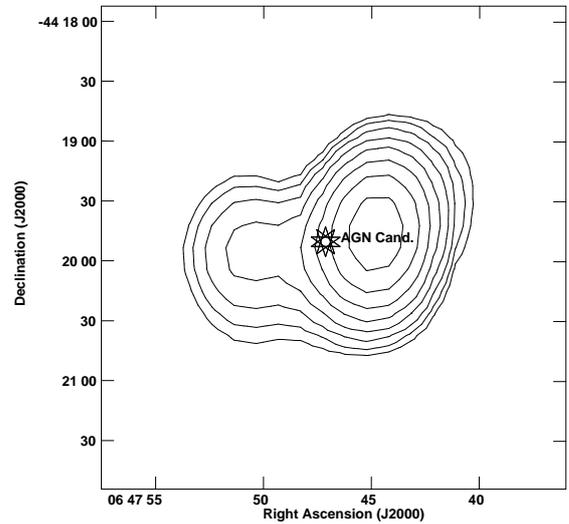}
\end{center}
\vspace{-2.0cm}
\caption{Zoomed countour map of SUMSS radio emission at 843 MHz towards the X-ray  candidate AGN 3XMM J064747.1$-$441950, whose position is marked by the star symbol.
Contours show start at four times the rms noise of 1.3 mJy/beam and proceed in $\sqrt{2}$ steps.
\label{contour}
 }
\end{figure}

\begin{figure}
\begin{center}
\includegraphics[scale=0.25]{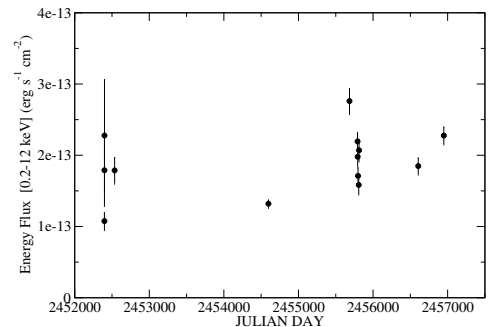}
\end{center}
\caption{Archival X-ray light curve of the candidate AGN 3XMM J064747.1$-$441950 in the energy range 0.2-12 keV.
\label{lcxrays}
 }
\end{figure}

\begin{figure}
\begin{center}
\includegraphics[scale=0.25, angle=-90]{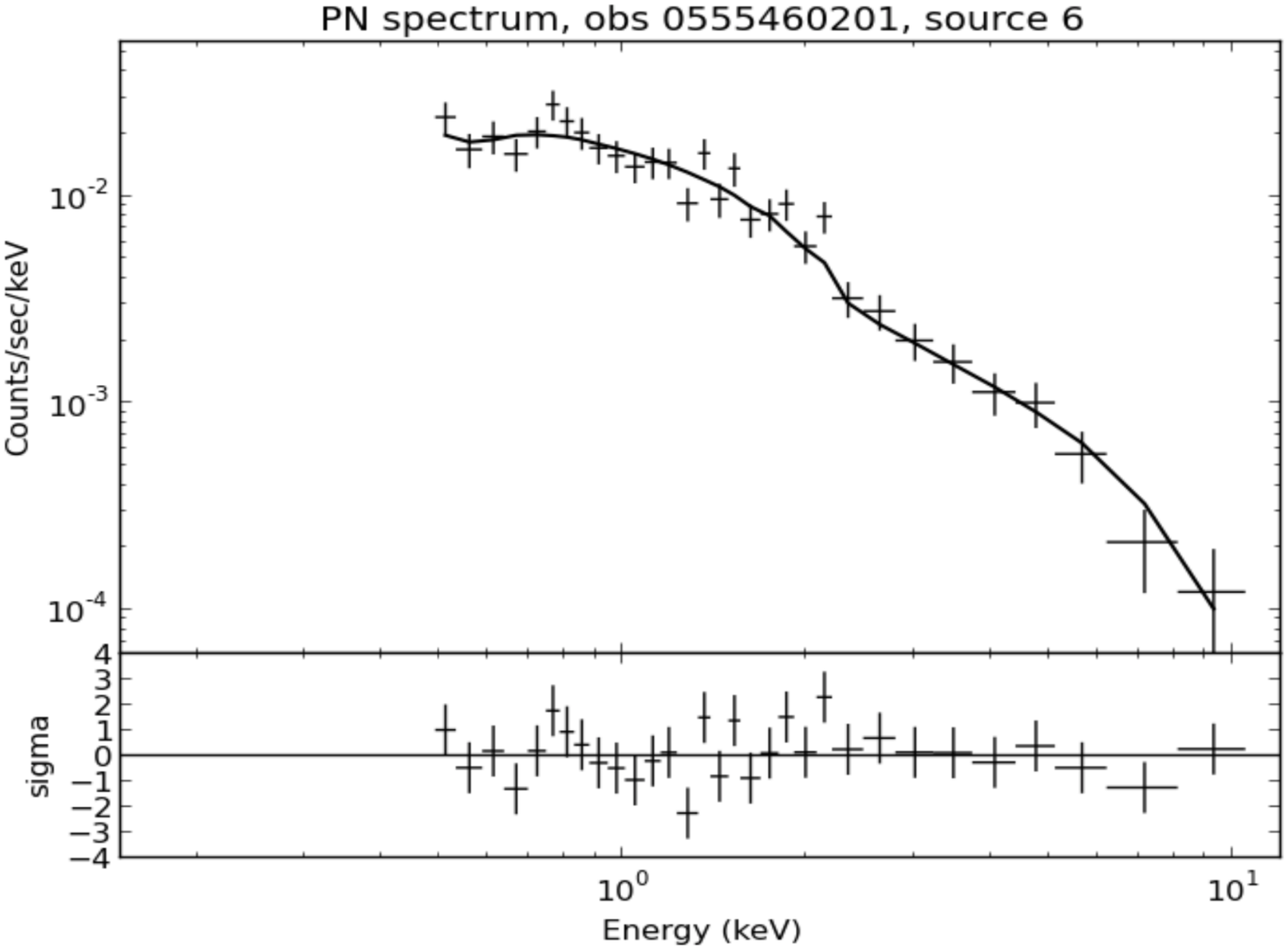}
\end{center}
\caption{Archival spectrum of the candidate AGN 3XMM J064747.1$-$441950 as observed with the EPIC pn cameras on board the
XMM-{\it Newton} observatory during a 42 ks exposure (date 5th October 2008).
\label{xrays}
 }
\end{figure}

At first glance, the candidate AGN 3XMM J064747.1$-$441950 within the 4FGL J0647.7$-$4418 location does not seem to match any of the radio sources
in the field according to the SUMSS (red layer in Fig. \ref{tricom}). 
We also checked this using the 150 MHz data from the TIFR GMRT Sky Survey (TGSS), 
with slightly better angular resolution but less sensitivity,
as provided by its first alternative data release \citep{2017A&A...598A..78I}.
The closest SUMSS radio source is coincident with TGSS  J064744.7$-$441947, offset by about $24^{\prime\prime}$ approximately to the West. 
It is worth to mention here the existence of  a second fainter SUMSS detection, absent in the TGSS, located about $40^{\prime\prime}$ away to the East
and with a roughly opposite position angle with respect to the proposed AGN. 
This relative position is morphologically reminiscent of a pair of  extended radio lobes as displayed in the Fig. \ref{contour} zoomed contour plot. 
Despite the poor angular resolution of the survey, the $\sim4$ brightness contrast between the two apparent lobes is suggestive of possible Doppler-boosted jets
emanating from 3XMM J064747.1$-$441950, which is located just in between. Moreover, 
it is worth to mention that
when the SUMSS flux density of the brightest radio source is combined with its TGSS measurement, a tentative non-thermal spectral index 
$\alpha \simeq -0.6 \pm 0.3$ is obtained assuming constant emission.

Retrieving flux measurements from the XMM-{\it Newton} Survey Science Centre\footnote{http://xmm-catalog.irap.omp.eu}, 
3XMM J064747.1$-$441950 appears as a persistent  X-ray source over time scales of years.
In Fig. \ref{lcxrays}, we compiled the available X-ray light curve in the 0.2-12 keV range. 
The source is clearly variable by a factor of $\sim2$ on time scales of at least months.
Unfortunately, its time sampling is not dense enough for a meaningful comparison
with {\it Fermi} LAT data. When observed with the longest on-source time (see Fig. \ref{xrays}), the source X-ray spectrum was better fitted by a simple absorbed power-law
with photon index $1.8\pm 0.1$ and hydrogen column density $(8\pm1) \times 10^{20}$ cm$^{-2}$. The corresponding reduced $\chi^2$ value amounted to 1.1 over 28 degrees of freedom.

\subsection{Photometric support for a blazar identification}

\begin{figure}
\begin{center}
\includegraphics[scale=0.4]{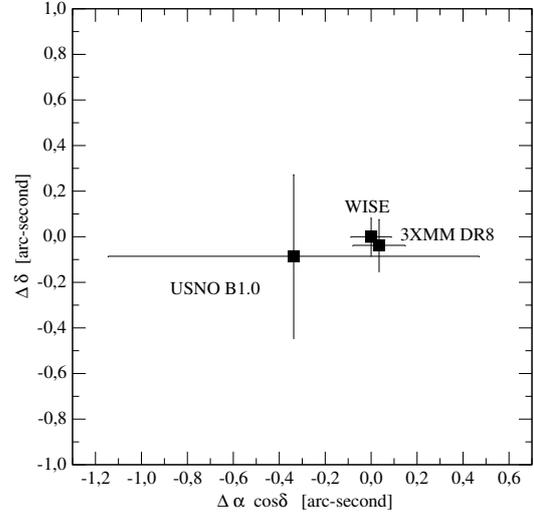}
\end{center}
\vspace{-0.5cm}
\caption{Relative positions of the X-ray source 3XMM J064747.1$-$441950, the optical source USNO\_B1\_0 \# 0456-0065720, and the 
infrared source AllWISE J064747.12-441950.3, which we propose to originate from  the same object.
\label{astromet}
 }
\end{figure}

Further inspection of POSS II plates additionally reveals a faint  optical source whose
 position is in sub-arsecond agreement with  the 'central' X-ray source 3XMM J064747.1$-$441950. Shining at  $R\simeq 19.0$, this point-like object corresponds to entry
\# 0456-0065720 in the USNO\_B1\_0 Catalog   \citep{2003AJ....125..984M}, where its lack of proper motion also agrees with an extragalactic origin.
In the absence of spectroscopic information, a reassuring information about the nature of this AGN candidate comes from the infrared domain.
This is  thanks to the Wide-field Infrared Survey Explorer (WISE) and its associated catalogue releases  \citep{2010AJ....140.1868W, 2013yCat.2328....0C}, 
where we find the AllWISE J064747.12-441950.3 source.
To confirm as confidently as possible that the XMM, USNO and this WISE object are actually the same, we plot their relative positions in Fig. \ref{astromet} where their mutual match agrees
within a fraction of an arc-second.

\begin{figure}
\begin{center}
\includegraphics[scale=0.4]{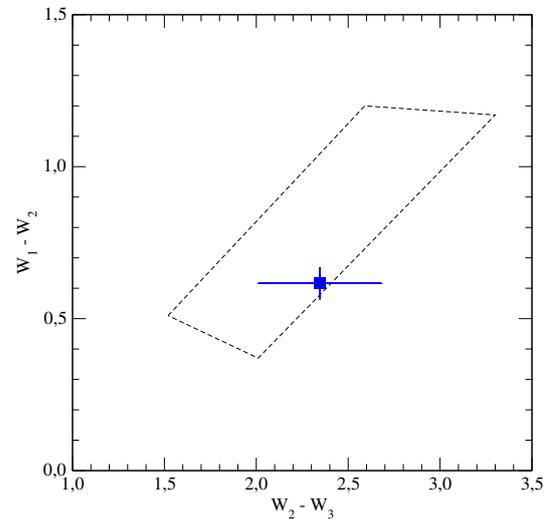}
\end{center}
\vspace{-0.8cm}
\caption{Colour-colour diagram showing the WISE Blazar Strip limits  as dashed lines and the location of the WISE source J064747.12$-$441950.3 in blue.
\label{strip}
 }
\end{figure}

The WISE catalogue series enabled the development of a powerful photometric tool to identify blazar AGNs within the error box of {\it Fermi} unassociated sources.
We refer here to the WISE Blazar Strip technique developed some years ago and successfully put to test
in recent times \citep{2012ApJ...750..138M, 2016ApJ...827...67M}. These authors discovered that non-thermal
emission in blazars clusters them in a narrow strip of the WISE colour-colour diagram, thus significantly facilitating their identification.

In our case, the WISE source magnitudes
in the 3.4 $\mu$m, 4.6 $\mu$m and 11.6 $\mu$m relevant bands are $W_1=15.445\pm0.037$, $W_2=14.828\pm0.052$ and $W_3=12.482\pm 0.331$, respectively.
The corresponding colours have been plotted in Fig. \ref{strip} together with the accepted WISE Blazar Strip limits. The agreement is remarkable and we believe this has to be taken
as an indication that \fgl\ is more likely to be a blazar than associated with the X-ray binary HD 49798.
Moreover, in Fig. 5 of \citet{2016ApJ...827...67M},   a strong correlation between the {\it Fermi} spectral index and the $W_1 - W_2$ colour was also established for blazars.
Plotting our source onto it (not shown here) gives further support to the proposed blazar.

\subsection{A blazar-like spectral energy distribution}

\begin{figure}
\begin{center}
\includegraphics[scale=0.34]{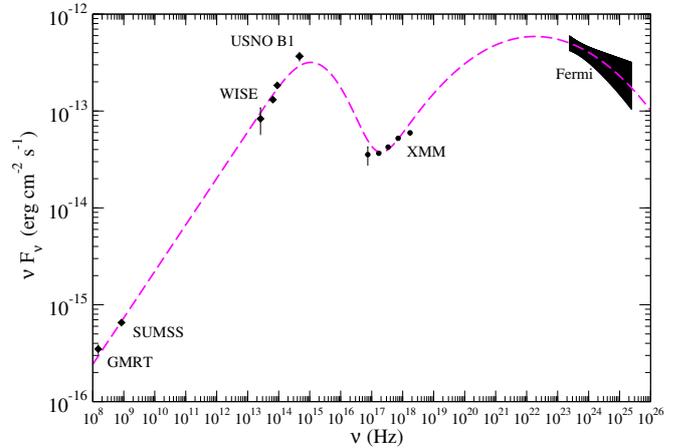}
\end{center}
\caption{Tentative SED of the proposed blazar source 4FGL J0647.7$-$4418. The magenta dashed line is a fitting attempt using the analytic expression by
\citet{2017MNRAS.468.4902P} and the parameters quoted in the text.
\label{sed}
 }
\end{figure}

In an attempt to further characterize the nature of our blazar candidate, we have assembled in Figure \ref{sed} a tentative spectral energy distribution (SED).
This plot combines the multi-wavelength data from the counterparts attributed in previous sections to 4FGL J0647.7$-$4418 
 at lower energies (radio, infrared, optical and X-rays).
 The Galactic absorption correction was estimated using NASA's HEASARC tool for hydrogen column density estimates\footnote{https://heasarc.gsfc.nasa.gov/cgi-bin/Tools/w3nh/w3nh.pl}.
This tool provided an average value of $(6\pm2) \times 10^{20}$ cm$^{-2}$  within 0.5 degrees around the target. We used this number, together with the 
\citet{1983ApJ...270..119M} cross sections, to place X-ray flux measurements in the XMM-{\it Newton} bands into the SED plot in terms of unabsorbed  flux per unit of frequency and
times the frequency. The corresponding optical extinction turns out to be very low ($A_R \simeq 0.2$ mag)
and practically negligible at infrared wavelengths.

 The resulting SED overall appearance is strongly reminiscent of blazar sources. The available data points, although scarce, clearly outline the existence of the two 
 peaks of synchrotron and inverse Compton origin typically displayed by blazar SEDs. To narrow the possible type of blazar powering the
 4FGL J0647.7$-$4418 emission, we tried to fit the SED points using the \citet{2017MNRAS.468.4902P} approach to unveil blazars among multi-wavelength counterparts
 of {\it Fermi} unassociated sources. Their parametrised SED fits use an analytic expression for the blazar luminosity $L_{\nu}$.
 We adapt it here to the monochromatic flux $F_{\nu}$ as a function of frequency $\nu$ as follows:
 \begin{equation}
 \nu F_{\nu} = \nu F_1(\nu) + \nu F_2 (\nu),   \label{sedfit}
 \end{equation}
 where
 \begin{equation}
 \nu F_i (\nu) = A_i \left(\frac{\nu}{\nu_i}\right)^{1-\alpha} \exp{\left\{-\frac{1}{2\sigma_i^2}  \left[  \log{\left(1+ \frac{\nu}{\nu_i} \right)}     \right]^2     \right\}}
 \end{equation}
 represents the synchrotron ($i=1$) and inverse Compton ($i=2$) peaks, respectively. The seven parameters of Eq. \ref{sedfit} include 
 $A_1$ and $A_2$ for normalization, the two frequency bumps $\nu_1$ and $\nu_2$, their widths $\sigma_1$ and $\sigma_2$, and a common spectral index $\alpha$.
 
 Unfortunately, we cannot introduce any redshift correction because spectroscopic observations are not available and we are forced to proceed assumig $z=0$.
 Moreover, the SED data available provides an insufficient sampling of the frequency space to enable the convergence of a full simultaneous least squares fit.
 To mitigate this added problem, we started by estimating the spectral index with a simple power-law fit restricted to the low-frequency region of radio and infrared data.
 The result was $\alpha=0.52 \pm 0.01$ and this parameter was kept fixed at all times. Later we proceeded to solve iteratively for each of the rest of parameters
 under a least squares criterion.
 The most plausible fit was obtained for
 $A_1=(2.1 \pm 0.6) \times 10^{-13}$ erg s$^{-1}$ cm$^{-2}$,
 $A_2=(5.2 \pm 0.2) \times 10^{-14}$ erg s$^{-1}$ cm$^{-2}$,
  $\sigma_1= 0.9 \pm 0.1$,  $\sigma_2=2.0 \pm 0.2$,
  $\nu_1=(1.3 \pm 0.6) \times 10^{14}$ Hz, and $\nu_2=(7.6 \pm 0.3) \times 10^{17}$ Hz.
  
 Despite the SED limited sampling, both in frequency and time, our best guess for the $\nu_1$ parameter suggests that we are dealing with
 a Low-Synchrotron-Peaked (LSP), or perhaps an Intermediate-Synchrotron-Peaked (ISP), blazar. The distinction between the two classes is arbitrarily set for
 $\nu_1$ below or above  $10^{14}$  Hz \citep{2015ApJ...810...14A} and we are just in the transition region.
 
 The classification of blazars is also possible using flux ratios, or broad band spectral indices. From our SED analysis, we find the following approximate
 flux ratios between radio, X-rays and gamma-rays: $\log{(F_{\rm radio}/F_{\rm X})} \simeq 6.9$ and $\log{(F_{\rm gamma}/F_{\rm X})} \simeq -5.8$.
 The corresponding spectral indices are $\alpha_{\rm radio-X}\simeq 0.80$ and $\alpha_{\rm X-gamma} \simeq 0.87$,
 with the reference frequencies being set to 1.4 GHz, 2 keV and 10 GeV.
 Bringing these values into the classification plots of \citet{2017MNRAS.468.4902P} Fig. 26 (not shown here), they appear to be also located in the region in between LSP and ISP blazars, thus supporting our preliminary classification.

\section{Conclusions}

The main conclusion of this work is a strong word of caution concerning the association of the gamma-ray source
\fgl\ with the O-type X-ray binary HD 49798 as proposed in the latest {\it Fermi} catalogue.

In addition, an alternative  counterpart candidate of likely blazar nature, the X-ray source 3XMM J064747.1$-$441950, almost centered inside the
{\it Fermi} LAT 95\% confidence ellipse, is proposed as a more reliable identification.
This statement is mainly supported by:
 i) the existence of non-thermal radio emission symmetrically located around it (SUMSS and GMRT sources);
 ii) the non-detection of the binary orbital period in the {\it Fermi} LAT light curve;
  iii) the agreement of the object's infrared colours with the well established WISE Blazar Strip; 
 iv) the similarity of the double-peaked SED with that of LSP/ISP blazars;
 v) the agreement  of radio/X-ray/gamma flux ratios and spectral indices also with LSP/ISP blazars.
Taken together, all  these facts would render \fgl\ a very ordinary blazar emitter of gamma rays instead of a new addition to the scarce family of GBs.

\section*{Acknowledgements}

This work was supported by the Agencia Estatal de Investigaci\'on grants  AYA2016-76012-C3-1-P and 
AYA2016-76012-C3-3-P from the Spanish Ministerio de Econom\'{\i}a y Competitividad (MINECO), 
by Consejer\'{\i}a de Econom\'{\i}a, Innovaci\'on, Ciencia y Empleo of Junta de Andaluc\'{\i}a under research group FQM-322, 
by grant MDM-2014-0369 of the ICCUB (Unidad de Excelencia 'Mar\'{\i}a de Maeztu'), and by the Catalan DEC grant 2017 SGR 643,
as well as FEDER funds. RC acknowledges support from NASA {\it Fermi} grant NNX15AU83G.




\bibliographystyle{mnras}
\bibliography{references} 








\bsp	
\label{lastpage}
\end{document}